\documentstyle[12pt]{article}
\newcommand{\beq}{\begin{equation}}
\newcommand{\eeq}{\end{equation}}
\newcommand{\beqn}{\begin{eqnarray}}
\newcommand{\eeqn}{\end{eqnarray}}

\newcommand{\p}{\mbox{${\vec p}$}}
\newcommand{\q}{\mbox{${\vec q}$}}
\newcommand{\r}{\mbox{${\vec r}$}}
\newcommand{\x}{\mbox{${\vec x}$}}
\newcommand{\n}{\mbox{${\vec n}$}}
\newcommand{\bk}{\mbox{${\vec k}$}}
\newcommand{\bq}{\mbox{${\vec q}$}}
\newcommand{\bv}{\mbox{${\vec v}$}}
\newcommand{\s}{\mbox{${\vec s}$}}
\newcommand{\bS}{\mbox{${\vec S}$}}
\newcommand{\si}{\mbox{${\vec \sigma}$}}
\newcommand{\Si}{\mbox{${\vec \Sigma}$}}

\newcommand{\Alpha}{\mbox{${\vec \alpha}$}}
\newcommand{\vrho}{\mbox{${\vec \rho}$}}
\newcommand{\vzeta}{\mbox{${\vec \zeta}$}}
\newcommand{\vom}{\mbox{${\vec \omega}$}}
\newcommand{\vOm}{\mbox{${\vec \Omega}$}}
\newcommand{\bu}{\mbox{${\vec u}$}}
\newcommand{\E}{\mbox{${\vec E}$}}
\newcommand{\B}{\mbox{${\vec B}$}}

\newcommand{\f}{\mbox{${\vec f}$}}
\newcommand{\fib}{\mbox{${\vec \phi}$}}
\newcommand{\va}{\mbox{${\vec a}$}}
\newcommand{\ep}{\mbox{${\epsilon}$}}
                 
\begin{document}

\begin{titlepage}

\begin{center}
{\bf EQUATIONS OF MOTION\\ OF SPINNING RELATIVISTIC PARTICLE\\
IN EXTERNAL FIELDS}\footnote{Plenary talk at the Fairbank 
Meeting on the Lense--Thirring Effect, Rome-Pescara, 29/6-4/7 
1998}\\
\vspace{0.5cm}
{\bf I.B. Khriplovich and A.A. Pomeransky}\\
\vspace{0.5cm}
Budker Institute of Nuclear Physics, 
630090 Novosibirsk, Russia\\
and Novosibirsk University\\
E-mail: khriplovich@inp.nsk.su, pomeransky@vxinpz.inp.nsk.su
\end{center}    

\vspace{1.0cm}

\begin{abstract}
We consider the motion of a spinning relativistic particle in
external electromagnetic and gravitational fields, to first order
in the external field, but to an arbitrary order in spin.
The noncovariant spin formalism is crucial for the correct 
description of the influence of the spin on the particle 
trajectory.  We show that the true coordinate of a relativistic 
spinning particle is its naive, common coordinate $\r$.
Concrete calculations are performed up to second order in spin
included. A simple derivation is presented for the gravitational
spin-orbit and spin-spin interactions of a relativistic particle.
We discuss the gravimagnetic moment (GM), a specific spin effect
in general relativity. It is shown that for the Kerr black
hole the gravimagnetic ratio, i.e., the coefficient at the GM,
equals unity (just as for the charged Kerr hole the
gyromagnetic ratio equals two). The equations of motion obtained
for relativistic spinning particle in external gravitational field
differ essentially from the Papapetrou equations.

\end{abstract}

\end{titlepage}

\section{Introduction}
The problem of the motion of a particle with internal angular 
momentum
(spin) in an external field consists of two parts: the description
of the spin precession and accounting for the spin influence on
the trajectory of motion. To lowest nonvanishing order in $c^{-2}$
the complete solution for the case of an external electromagnetic
field was given more than 70 years ago~\cite{tho}. The gyroscope
precession in a centrally symmetric gravitational field had been
considered to the same approximation even earlier~\cite{des}. Then,
much later the spin precession was investigated in the case
of the gravitational spin-spin interaction~\cite{sch}. The fully
relativistic problem of the spin precession in an external
electromagnetic field was also solved more than 70 years 
ago~\cite{fr}, and then in a more convenient formalism, using the 
covariant vector of spin, in~\cite{bmt}.

The situation with the second part of the problem, which refers to
how the spin influences the trajectory, is different. Covariant
equations of motion for a relativistic spinning particle in an
electromagnetic field were written in the same paper~\cite{fr}, and
for the case of a gravitational field in~\cite{pa}. These
equations have been discussed repeatedly from various points of
view in numerous papers (see, e.g.,~[7-18]).
The problem of the influence of the spin on the trajectory of a 
particle in
external fields is not only of purely theoretical interest. It
attracts attention being related to the description of the motion
of ultrarelativistic particles in accelerators~\cite{dk} (see also
recent review~\cite{hei}).

In fact, it is far from being obvious whether one can observe
in practice the discussed spin corrections to the equations of
motion of elementary particles, say, electron or proton.
According to the well-known argument by Bohr (see~\cite{pau}), 
an additional Lorentz force due to the finite
size of the wave packet of a charged particle and to the
uncertainty relation, exceeds the corresponding component
of the Stern--Gerlach force. However, this argument by itself 
does not exclude in principle the possibility to observe a
regular Stern--Gerlach effect, let even a small one, in the 
presence of a larger background due to the uncertainty relation.
This possibility is supported, in particular, by recent 
calculations~\cite{bgs}. Moreover, spin-dependent correlations 
certainly exist in
differential cross sections of scattering processes. So, it was 
proposed long ago to separate by polarizations a beam of charged
particles in a storage ring through the spin interaction
with external fields~\cite{ros}. Though this proposal is
being discussed rather actively (see review~\cite{hei}),
it is not clear up to now whether it is feasible technically.

There are however macroscopic objects for which internal rotation 
certainly influences their trajectories. We mean the motion of
Kerr black holes in external gravitational fields. This problem is 
of
importance in particular for the calculation of the gravitational
radiation of binary stars. In this connection it was considered 
in~[24-27]. However, when we turned to these
calculations, we found~\cite{khp} that the equations of motion
with spin taken into account to the lowest nonvanishing order in
$c^{-2}$, used in these papers, even in the simpler case of an
external field lead to results which differ from the well-known
gravitational spin-orbit interaction. The problem is essentially
related to the correct definition of the center-of-mass coordinate.
Moreover, it turned out that the widely used Papapetrou 
equations~\cite{pa} 
also fail to reproduce in the same $c^{-2}$ approximation the 
result for the gravitational spin-orbit interaction found in the
classical work~\cite{des}. This discrepancy was pointed out long ago 
in~\cite{bar}; however the explanation suggested in~\cite{bar} does 
not look satisfactory (see~\cite{khp}).

In our recent work~\cite{khp1} (the present talk is essentially 
based on it) we derived equations of motion of a relativistic 
particle with the noncovariant description of spin. These equations 
agree with well-known limiting cases. Though for external
electromagnetic field such equations in the linear in spin
approximation have been obtained previously~\cite{dk} (see 
also~\cite{hei}), we would like to start with comments related to 
this approximation in electrodynamics.

\section{Covariant and Noncovariant Equations\\ of Motion
of a Spinning Particle\\ in an Electromagnetic Field}
\subsection{The Problem with Covariant Equations of Motion}
The interaction of spin with external electromagnetic field
is described, up to terms on the order of $c^{-2}$ included,
by the well-known Hamiltonian (see, for example,~\cite{blp})
\begin{equation}\label{th}
H=-\,\frac{eg}{2m}\s\B+\,
\frac{e(g-1)}{2m^2}\s[\p \times \E] \,.
\end{equation}
Here $\B$ ¨ $\E$ are external magnetic and electric fields;
$e$, $m$, $\s$, and $\p$ are the particle charge, mass, spin, and
momentum, respectively; $g$ is its gyromagnetic ratio.
Let us emphasize that the structure of the second (Thomas)
term in this expression not only has been firmly established
theoretically, but has also been confirmed with high accuracy
experimentally, at any rate in atomic physics. To avoid 
misunderstandings, let us note that, generally speaking, 
the last term in formula (\ref{th}) should be rewritten in a 
Hermitian form (see, e.g.,~\cite{bjd}):
\[ [\p \times \E ] \rightarrow {1 \over 2}\,
\left([\p \times \E ] - [\E \times \p ]\right) 
= [\p \times \E ] + {i \over 2}\,\nabla \times \E \,. \] 
We will be interested, however, in the semiclassical approximation 
mainly, when in the interaction linear in spin, field derivatives 
are neglected. (Besides, the correction with $\nabla \times \E$
vanishes in the case of potential electric field considered
in~\cite{blp}.) 

Let us try to construct a covariant equation of motion
accounting for spin, which would reproduce in the same
approximation the force  
\begin{equation}\label{thac}
\f_m=\,\,\frac{eg}{2m}\s\B,_m+\,\frac{e(g-1)}{2m}
\,\left(\frac{d}{dt}[\E \times \s\,]_m-\,
\s[\bv \times \E,_m\,]\right),
\end{equation}
corresponding to the Hamiltonian (\ref{th}) (here and below coma 
with a subscript denotes a partial derivative). A covariant correction
$f^{\mu}$ to the Lorentz force $eF^{\mu\nu}u_{\nu}$, linear in
the tensor of spin $S_{\mu\nu}$ and in the gradient of the
tensor of electromagnetic field $F_{\mu\nu,\lambda}\,$, may
depend also on the 4-velocity $u^\mu$. Since
$u^{\mu}u_{\mu}=1$, this correction must satisfy the
condition $u_{\mu}f^{\mu}=0$. From the mentioned tensors one
can construct only two independent structures meeting
the last condition. The first,
\begin{equation}\label{co1}
\eta^{\mu\kappa}F_{\nu\lambda,\kappa}S^{\nu\lambda}\,-\,
F_{\lambda\nu,\kappa}u^{\kappa}S^{\lambda\nu}u^{\mu},
\end{equation}
reduces in the $c^{-2}$ approximation to
$$2\s(\B,_m-\,[\bv \times \E,_m]),$$
and the second,
\begin{equation}\label{co2}
u^{\lambda}F_{\lambda\nu,\kappa}u^{\kappa}S^{\nu\mu},
\end{equation}
reduces to
$$\frac{d}{dt}[\s \times \E]\,_m\,.$$
(Let us note that the structures with the contraction
$F_{\nu\kappa,\lambda}S^{\kappa\lambda}$ reduce to these two
expressions, due to the Maxwell equations and the
antisymmetry of $S_{\kappa\lambda}$.) 

Obviously, no linear combination of these two structures 
can reproduce the correct expression (\ref{thac}) for the 
spin-dependent force. In a somewhat less general way it was 
shown in~\cite{khp}. 

But why the correct (in the $c^{-2}$ approximation) formula
(\ref{thac}) cannot be obtained from a covariant expression for the 
force? Obviously, one can easily reproduce those terms in 
(\ref{thac}) which are proportional to $g$ by a linear combination 
of (\ref{co1}) and (\ref{co2}), i.e., there is no problem to 
present in a covariant form the 
terms which describe, so to say, direct interaction 
of a magnetic moment with external fields. It is the terms in 
(\ref{thac}) independent of $g$ and corresponding to the Thomas 
precession which cannot be written covariantly. Certainly, the 
Thomas precession can be described beyond the $c^{-2}$ 
approximation, for arbitrary velocities. But there are no 
reasons why this essentially noncovariant phenomenon should have a 
covariant description. This is the point.

\subsection{What Is the Correct Definition of the Coordinate
of a Spinning Particle?}
It was noted in~\cite{khp} that the covariant formalism
can be reconciled with the correct results if the coordinate $\x$ 
entering the covariant equation is related to the usual one $\r$ in 
the $c^{-2}$ approximation as follows:
\begin{equation}\label{def}
\x\,=\,\r+\,\frac{1}{2 m}\s\times\bv. 
\end{equation}.  
The generalization of this substitution to the case of arbitrary 
velocities 
\begin{equation}\label{defh}
\x\,=\,\r+\,\frac{\gamma}{m(\gamma+1)}\s\times\bv,\;\;\;
\gamma =\,{1 \over \sqrt{1-v^2}}. 
\end{equation}
was recently pointed out in~\cite{hei}.

But why the spin precession itself (as distinct from
the spin influence on the trajectory) can be described 
covarianly~\cite{fr,bmt} without any concern for the coordinate
definition? First of all, the covariant equations of the spin 
precession
\beq\label{bmt0}
\frac{dS_{\mu}}{d\tau}=\,\frac{e}{2m}\left[g F_{\mu\nu}S^{\nu}
-(g-2)u_{\mu}F_{\lambda\nu}u^{\lambda}S^{\nu}\right]
\eeq
(here $S_{\mu}$ is the covariant 4-vector of spin) are written in 
the semiclassical approximation, i.e., the coordinate dependence of 
external fields is neglected at all. Second, equations (\ref{bmt0}) 
are homogeneous and linear in spin. So, even if one 
went here beyond the semiclassical approach, but stayed within the 
approximation linear in spin, the use of the usual coordinate $\r$,
which differs from $\x$ in terms proportional to $\s$ only,
would be completely legitimate.

Of course, the choice of the variable, $\r$ or $\x$, is by itself
a matter of convention. But still, which of them is the true 
center-of-mass coordinate of a relativistic spinning body?

We note first of all that relation (\ref{def}) is valid for a free 
particle as well. So, to answer the question, we will 
consider a simple example of the free Dirac particle with the 
Hamiltonian
$$H\,=\,\Alpha \p\, + \,\beta\,m\,.$$ 
Here, the operator whose expectation value equals to $\r$, is not 
$\r$ itself, but~\cite{fw}
\begin{equation}\label{co}
\x\,=\,\r\,+\,\frac{i\beta \Alpha}{2 \ep}\,
-\,\frac{i\beta(\Alpha \p)\p\,+\,[\Si \times
\p]\,\ep}{2 \ep^2 (\ep+m)};\;\;\ep=\,\sqrt{p^2+m^2};\;\;
\Si\,=\,\frac{1}{2i}[\Alpha \times \Alpha].
\end{equation}
To lowest nonvanishing order in $c^{-2}$ expression (\ref{co}) 
reduces to
\begin{equation}\label{con}
\x\,=\,\r\,+\,\frac{1}{2 m}\,\s\times\bv\,,\;\;\;
\s\,=\,\frac{1}{2}\si\,,
\end{equation}
which might prompt indeed substitution (\ref{def}). However, 
under the Foldy--Wouthuysen (FW) transformation, which separates 
positive-energy states from negative-energy ones, the relativistic 
operator $\x$ goes over into mere $\r$. And the
transition from the exact Dirac equation in an external field to 
its approximate form containing only the first-order correction in 
$c^{-2}$ is performed just by means of the FW transformation. Thus, 
in the arising $c^{-2}$ Hamiltonian the coordinate of a spinning 
electron is the same $\r$ as in the completely nonrelativistic 
case. Nobody makes substitution (\ref{def}) in the Coulomb potential 
when treating the spin-orbit interaction in the hydrogen atom.

As to a classical particle, it is in fact a well-localized 
wave packet constructed from positive-energy states, i.e., it is 
properly described in the FW representation. Therefore, it 
is $\r$ which is the true coordinate of a classical relativistic 
spinning particle.

\subsection{The Noncovariant Formalism}
The correct equations of motion in an electromagnetic
field including spin to first order are known for a fairly
long time~\cite{dk}. Though being fully relativistic, they are
noncovariant and based on the initial physical definition of spin.
According to this definition, spin is the 3-dimensional vector $\s$ 
(or 3-dimensional antisymmetric tensor $s_{mn}$) of the internal 
angular momentum defined in the rest frame of the particle. The 
covariant vector of spin $S_{\mu}$ (or the covariant antisymmetric 
tensor $S_{\mu\nu}$) are obtained from $\s$ 
(or $s_{mn}$) merely by the Lorentz transformation. By the way, an 
advantage of this approach is that the constraints
$u^{\mu}S_{\mu}=0$ and $u^{\mu}S_{\mu\nu}=0$ hold identically.
The precession frequency for spin $\s$ at an arbitrary velocity
is well-known (see, for instance,~\cite{blp}):
$$ {\vOm}=\,\frac{e}{2m}\,\left\{(g-2)\,\left[\B
-\,\frac{\gamma}{\gamma+1}\,\bv(\bv\B)\, -\,\bv \times \E
\right]\,\right.$$
\beq\label{Om}
\left.+2\,\left[\frac{1}{\gamma}\,\B
-\,\frac{1}{\gamma+1}\,\bv \times \E\right]\right\}.
\eeq
Naturally, the corresponding
interaction Lagrangian (the Lagrangian description is here
somewhat more convenient than the Hamiltonian one) equals
\[ L_{1s} = {\vOm}\s=\,\frac{e}{2m}\,\s\,
\left\{(g-2)\,\left[\B
-\,\frac{\gamma}{\gamma+1}\,\bv(\bv\B)\, -\,\bv \times \E
\right]\,\right. \]
\beq\label{lse}
\left.+2\,\left[\frac{1}{\gamma}\,\B
-\,\frac{1}{\gamma+1}\,\bv \times \E\right]\right\}.
\eeq
The equation of motion for coordinate is the usual one:
\beq\label{fs}
(\nabla - \frac{d}{dt} \nabla_{\bv})L_{tot}=0,
\eeq
where $L_{tot}$ is the total Lagrangian of the system.
The equation of motion for spin in general form is
\beq\label{dsc}
\dot{\s} = - \{L_{tot},\s\},
\eeq
where $\{...\;,...\}$ is the Poisson bracket, or
\beq\label{dsq}
\dot{\s} = - i[L_{tot},\s]
\eeq
in the quantum problem. This applicability of a common canonical 
formalism is one more advantage of the noncovariant 
approach. Meanwhile, in the covariant approach we have to deal 
in particular with higher time derivatives, which is obvious 
already from relation (\ref{con}).

\section{Equations of Motion of a Spinning Particle\\
         in an Electromagnetic Field}
\subsection{The General Approach}
Our approach to the problem is based on the following physically
obvious argument. As long as we do not consider excitations of
internal degrees of freedom of a body moving in an external field, 
this body (even if it is macroscopic) can be treated as an
elementary particle with spin. 

Therefore, the Lagrangian of the spin interaction with an external 
field can be derived from the elastic scattering amplitude
\beq\label{ampl}
- e J^{\mu}A_{\mu}
\eeq
of a particle with spin $s$ by a vector potential $A_{\mu}$. Due to 
the arguments presented in Section 1, the discussion of the effects
nonlinear in spin (which are of primary interest to us) may be 
physically meaningful only in the classical limit $s\gg 1$. It is 
this approximation that is basically used below.

The matrix element $J_{\mu}$ of the electromagnetic
current operator between states with momenta $k$ and
$k^\prime$ can be written (under $P$ and $T$ invariance)
as follows (see, for instance,~\cite{khri,kms}):
\beq\label{cur}
J_{\mu}=\frac{1}{2\ep}\,\bar{\psi}(k^\prime)
\left\{p_{\mu}F_e\right.
\left. +\, \Sigma_{\mu\nu}q^{\nu}\,F_m\right\}\,\psi(k).
\eeq
Here $\;p_{\mu}=(k^\prime+k)_{\mu},
\;\; q_{\mu}=(k^\prime-k)_{\mu}$.

The wave function of a particle with an arbitrary spin
$\psi$ can be written (see, for instance,~\cite{blp}, \S 31) as
\begin{equation}\label{ps}
            \psi={1 \over \sqrt 2} \left( \begin{array}{c} \xi\\
                                         \eta\\
                        \end{array}
                 \right).
\end{equation}
Both spinors,
\[ \xi=\{ \xi^{{\alpha}_1\,{\alpha}_2\,\,..\,\,
{\alpha}_p\,}_{\dot{\beta}_1
\,\dot{\beta}_2\,..\,\,\dot{\beta}_q}\} \]
and
\[ \eta=\{ \eta_{\dot{\alpha}_1\,\dot{\alpha}_2
\,\,..\,\,\dot{\alpha}_p\,}^{{\beta}_1\,{\beta}_2\,..\,\,
{\beta}_q}\},
\]
are symmetric in the dotted and undotted indices separately,
and
\[ p+q=2s. \]
For a particle of half-integer spin one can choose
\[ p=s+\,\frac{1}{2}\,,\;\;\;\;q=s-\,\frac{1}{2}\,\,. \]
In the case of integer spin it is convenient to use
\[ p=q=s. \]
The spinors $\,\xi\,$ and $\,\eta\,$ are chosen in such a way
that under reflection they transform into each other (to within a
phase). For $p \neq q$ they are different objects which
belong to different representations of the Lorentz group.
For $p=q$, these two spinors coincide. Nevertheless, we will
use the same expression (\ref{ps}) for the wave function of
any spin, i.e., we will also formally introduce the object
$\,\eta\,$ for an integer spin, bearing in mind that it is
expressed in terms of $\,\xi\,$. This will allow us to
perform calculations in the same way for the integer and
half-integer spins.

In the rest frame both $\xi$ and $\eta$ coincide with a
nonrelativistic spinor $\xi_0$, which is symmetric in all
indices; in this frame there is no difference between dotted
and undotted indices. The spinors $\xi$ and $\eta$ are
obtained from $\xi_0$ through the Lorentz transformation:
\beq\label{lt}
\xi=\exp\{\Si\fib/2\} \xi_0\,;\;\;\;\;\;\;\;
\eta=\exp\{-\Si\fib/2\} \xi_0\,.
\eeq
Here the vector $\fib$ is directed along the velocity,
$\;\;\tanh\phi=v$;
\[  \Si \,=\,\sum_{i=1}^{p} \si_i\,-\,
\sum_{i=p+1}^{p+q} \si_i\,, \]
and $\si_i$ acts on the $i$th index of the spinor $\,\xi_0\,$
as follows:
\begin{equation}\label{aa}
\si_i\,\xi_0
=(\si_i)_{\alpha_i\beta_i}\, (\xi_0)_{....\beta_i...}\;.
\end{equation}
In the Lorentz transformation (\ref{lt}) for $\xi$, after
the operator $\Si$ has acted on $\xi_0$ the first $p$
indices are identified with the upper undotted indices and
the next $q$ indices are identified with the lower dotted
indices. The inverse situation takes place for $\eta$.

We note that in an external field the components of
velocity $\bv$ (and together with them the components of
$\fib$) do not commute, in general. However, to the adopted
approximation, linear in the external field, one can ignore 
this noncommutativity which is itself
proportional to the field. Moreover, we are mainly interested
in the classical limit of the final result
where such commutators are negligible since they are
proportional to an extra power of $\hbar$. Therefore,
$\bv$ and $\fib$ will be treated as ordinary numerical parameters.

Next,
\[ \bar{\psi} = \psi^\dagger \gamma_0 =
             \psi^\dagger  \left(
                            \begin{array}{rr}
                                0 & I \\
                                I & 0\\
                             \end{array}
                       \right);
      \]
here $I$ is the sum of $2\times 2$ unit matrices acting on
all indices of the spinors $\,\xi\,$ and $\,\eta\,$. The
components of the matrix $\Sigma_{\mu\nu}=-\Sigma_{\nu\mu}$
are:
\begin{equation}\label{0n}
\Sigma_{0n}= \left(
 \begin{array}{rr}
-\Sigma_n    & 0\;\; \\
0\;\; & \Sigma_n\\
\end{array}
                       \right);
      \end{equation}
\begin{equation}\label{mn}
      \Sigma_{mn}=\,-\,2i\ep_{mnk}\left(
                            \begin{array}{rr}
                                s_k & 0\;  \\
                                0\;   & s_k\\
                             \end{array}
                       \right);
      \end{equation}
\[ \s =\,\frac{1}{2}\sum_{i=1}^{2s} \si_i.  \]

The scalar operators $F_{e,m}$ depend on two invariants,
$t=q^2$ and $\tau=(S^{\mu}q_{\mu})^2$. The covariant vector
of spin $S_{\mu}$ is defined, e.g., for the state with
momentum $k_{\mu}$, and is obtained via the Lorentz
transformation from the vector of spin $(0,\s)$ in the rest
frame:
\beq\label{Ss}
S^{\mu}=(S_0,\bS),\;\;S_0=\frac{(\s\bk)}{m},\;\;
 \bS=\s+\frac{\bk(\bk\s)}{m(\ep+m)}.
\eeq
In the expansion in the electric multipoles
\[ F_e(t,\tau)=\sum_{n=0}^{N_e}f_{e,2n}(t)\tau^n \]
the highest power $N_e$ is obviously $s$ and
$s-1/2$ for integer and half-integer spin, respectively.
In the magnetic multipole expansion
\[ F_ m(t,\tau)=\sum_{n=0}^{N_m}f_{m,2n}(t)\tau^n \]
the highest power $N_m$ is $s-1$ and $s-1/2$ for
integer and half-integer spin. The invariant form factors 
$f_{e,m}(t)$, being Fourier-transformed into the coordinate 
representation, describe the space distribution of the charge, 
magnetic moment, and higher multipoles, i.e., describe the finite 
size of the particle.

Clearly,
$$f_{e,0}(0)=1,\;\;\;\;f_{m,0}(0)=\frac{g}{2}.$$

Let us note at last that we have chosen the noncovariant
normalization for the amplitude (\ref{ampl}), being
interested in the Lagrangian referring to the world time
$t$ and not to the proper time $\tau$.

\subsection{Effects Linear in the Spin}  
As a warm-up exercise and check of our approach, let us 
reproduce now the well-known result (\ref{lse}) for the case of a 
constant external field. We start with the terms proportional to 
$g$-factor. The corresponding term in the scattering amplitude can 
be written as
\[ \frac{eg}{4\ep}\;\xi_0^{\prime \dagger}
\left\{[\exp\{\Si\fib/2\}(\s\B)
\exp\{-\Si\fib/2\}+\exp\{-\Si\fib/2\}(\s\B)\exp\{\Si\fib/2\}]
\right. \]
\beq\label{lseg}
+\,{i \over 2}
\,[\exp\{\Si\fib/2\}(\Si\E)\exp\{-\Si\fib/2\}
\eeq
\[ 
\left.-\exp\{-\Si\fib/2\}(\Si\E)\exp\{\Si\fib/2\}]\right\}\xi_0.\]

It is essential that in the considered case of a
constant external field, one may put
$\;{\bk}^\prime={\bk},\;{\bv}^\prime
={\bv},\;\fib^\prime=\fib\,$, since $\;{\bq}
={\bk}^\prime-{\bk}\,$ corresponds to the field gradient.

In our further calculations we use the well-known identity
\[ \exp\{\hat A\} \hat B \exp\{-\hat A\}
= \hat B\,+\,\frac{1}{1!}\,[\hat A,\,\hat B]\,
+\,\frac{1}{2!}\,\left[\hat A,\,[\hat A,\,\hat B]\right]+...\; ,\]
and the following relationships:
\beq\label{ide1}
 [\Sigma_i,\,\Sigma_j]=4i\ep_{ijk}s_k\, ,\;\;\;
   [\Sigma_i,\,s_j]=i\ep_{ijk}\Sigma_k\, ;
\eeq
\beq\label{ide2}
\cosh \phi = \gamma,\;\;\; \sinh \phi = v\,\gamma.
\eeq
After simple algebraic transformations,
expression (\ref{lseg}) reduces to
\beq\label{lseg1}
\frac{eg}{2m}\,\s \left[\B
-\,\frac{\gamma}{\gamma+1}\,\bv(\bv\B)\, -\,\bv \times \E
\right].
\eeq

Let us now discuss the contribution of the convection term
\beq\label{conv}
-\,\frac{e}{2\ep}\,\bar{\psi}(k^\prime)\psi(k)\,p^{\mu}A_{\mu}.
\eeq
We write the product of exponents in the expression
\[ \bar{\psi}(k^\prime)\psi(k) \]
\beq\label{pro}
= {1 \over 2}\xi_0^{\prime \dagger}
[\exp\{\Si\fib^\prime/2\} \exp\{-\Si\fib/2\}
+\exp\{-\Si\fib^\prime/2\}\exp\{\Si\fib/2\}]\xi_0
\eeq
as  
\[ \exp\{\Si\fib^\prime/2\} \exp\{-\Si\fib/2\} \]
\beq\label{prod}
=\prod_p\exp\{\si\fib^\prime/2\} \exp\{-\si\fib/2\}
\prod_q\exp\{-\si\fib^\prime/2\} \exp\{\si\fib/2\}.
\eeq
Let us consider a typical factor in this formula:
\[ \exp\{\si\fib^\prime/2\} \exp\{-\si\fib/2\}
= \cosh(\phi^\prime/2)\cosh(\phi/2)
-(\n^\prime \n)\sinh(\phi^\prime/2)\sinh(\phi/2)\]
\beq\label{exp}
+\si\left[\n^\prime\sinh(\phi^\prime/2)\cosh(\phi/2)
-\n\cosh(\phi^\prime/2)\sinh(\phi/2)\right]
\eeq
\[-\,i\,(\si[\n^\prime \times \n])\sinh(\phi^\prime/2)\sinh(\phi/2);\]
here $\;\n^\prime=\bv^\prime/v^\prime,\;\n=\bv/v\,$. Since we
are interested in gradients only as long as they enter together
with spin, in the first term,
$\;\cosh(\phi^\prime/2)\cosh(\phi/2)
-(\n^\prime \n)\sinh(\phi^\prime/2)\sinh(\phi/2),\;$ we put
$\;\phi^\prime=\phi/2,\;\;\n^\prime = \n\, ,$ after which this
term turns to unity. Then, we are discussing  the interaction
linear in spin, so that the product (\ref{prod}) reduces to
\[ 1 + \Si\left[\n^\prime\sinh(\phi^\prime/2)\cosh(\phi/2)
-\n\cosh(\phi^\prime/2)\sinh(\phi/2)\right] \] \[
-\,2i\,(\s[\n^\prime \times \n])\sinh(\phi^\prime/2)\sinh(\phi/2).\]
When substituted into formula (\ref{pro}), the terms
proportional to $\Si$ cancel out. Now, limiting ourselves to the 
terms linear in $\q$, we reduce the spin-dependent part of
(\ref{conv}) to
\[ -\,e\,\frac{p^{\mu}}{2\ep}\,\frac{i(\s[\bk \times\q])}
{m(\ep+m)}\,A_{\mu}. \]
Let us note further that since $\;p^{\mu}q_{\mu}=0\,$, the
following identity holds
\beq\label{ide}
p^{\mu}q_{\alpha}A_{\mu}=p^{\mu}(q_{\alpha}A_{\mu}-q_{\mu}A_{\alpha})
=p^{\mu}iF_{\alpha\mu}.
\eeq
Then, we can put $\;p_{\mu} = 2 m u_{\mu}\,,$
where $\,u_{\mu}\,$ is the 4-velocity. As a result we arrive at
the following expression:
\beq\label{lse1}
-\,\frac{e}{2m}\,\s \left[2\,\left(1\,-\,{1\over \gamma}\right)\B
-\,\frac{2\gamma}{\gamma+1}\,\bv(\bv\B)\,
-\,\frac{2\gamma}{\gamma+1}\,\bv \times \E
\right].
\eeq
The sum of (\ref{lseg1}) and (\ref{lse1}) yields
(\ref{lse}). Thus, we have reproduced the well-known result for the
interaction linear in the spin, starting from the relativistic
wave equation for an arbitrary spin.

Below we repeatedly use identities of the form (\ref{ide}).
In the classical language such a transformation corresponds to
discarding in a Lagrangian (or adding to it) a total time 
derivative. Indeed,
\[ u^{\mu}q_{\mu}\rightarrow u^{\mu}\partial_{\mu}
=\,\gamma\,\left({\partial \over \partial t}\,+\,\bv \nabla\right)\,
=\,\gamma\,{d \over d t}. \]

\subsection{Effects Quadratic in the Spin}
Let us now investigate the interaction of second order in the spin.
The ``bare'', explicitly quadrupole interaction present in
expressions (\ref{ampl}) and (\ref{cur}) is
\beq\label{qu}
-\,e\,{p^{\mu} \over 2\ep}\,f_{e,2}(S^{\alpha}q_{\alpha})^2
\,A_{\mu}.
\eeq
Using the identity (\ref{ide}) and relations (\ref{Ss}), and then
discarding the total time derivative
$\partial/\partial t\,+\,\bv \nabla$,
we write this interaction as
\[ L_{2s}=\,-\,e\,f_{e,2}\,\left[(\s\nabla)\,-\,
{\gamma \over \gamma+1}\, (\bv\s)(\bv \nabla)\right] \]
\beq\label{qu1}
\times\left[(\s\E)\,
-\,{\gamma \over \gamma+1}\,(\s\bv)(\bv\E)\,
+\,(\s[\bv\times\B])\right].
\eeq

Using the Maxwell equations and adding a total derivative with 
respect to $t$, one can show that the tensor $s_i s_j$ in 
(\ref{qu1}) can be rewritten in the following irreducible form:
$s_i s_j\rightarrow s_i s_j-(1/3)\delta_{ij}\s^{\,2}$. Now from the 
nonrelativistic limit of formula (\ref{qu1}), it is clear that this 
formula describes indeed the interaction with an external field of
the quadrupole moment
\beq\label{qua}
Q_{ij}\,=\,-\,2\,e\,f_{e,2}\,(3\,s_i s_j-\,\delta_{ij}\s^{\,2});
\;\;\;Q\,=\,Q_{zz}\vert_{s_z=s}=\,-\,2\,e\,f_{e,2}\,s(2s-1).
\eeq
In the asymptotics, as $\gamma\rightarrow \infty$, the
interaction (\ref{qu1}) tends to a constant
\beq\label{quas}
L_{2s}=\,-\,e\,f_{e,2}\,[(\s\nabla)\,-\,(\bv\s)(\bv \nabla)]\,
[(\s\E)\,-\,(\s\bv)(\bv\E)\,+\,(\s[\bv\times\B])].
\eeq

It is well-known that even in the absence of the bare
quadrupole term, i.e., at $f_{e,2}=0$, a quadrupole interaction
arises in the nonrelativistic limit due to the convection and
magnetic terms in interaction (\ref{ampl}). The value of this
induced quadrupole moment at an arbitrary spin of the particle
is~\cite{kms} (in the formula below we corrected a misprint in 
the original paper~\cite{kms}):
\beqn\label{qua1}
Q_1=\,-\,e\,(g-1)\,\left({\hbar \over mc}\right)^2\,
\left\{\begin{array}{ll}
                               s, &  \mbox{integer spin,}\\
                               s-1/2,&\mbox{half-integer spin.}\\
                   \end{array}
             \right.
\eeqn
Here we have explicitly displayed the Planck
constant $\hbar$ to show that the induced quadrupole
moment $\;Q_1$ vanishes in the classical limit
$\;\;\;$$\hbar\rightarrow 0\;,$ $\;\;\;$ $\;s\rightarrow \infty,\; 
\hbar s\rightarrow$ const. Therefore,
the interaction of second order in spin proportional to $Q_1$
does not influence in fact equations of motion of a classical
particle (although it plays a role in atomic spectroscopy
\cite{kms}).

Still, the convection and magnetic terms in expression
(\ref{ampl}) induce an interaction of second order in spin
which has a classical limit and is therefore of interest for
our problem. It is convenient here to start with the convection
current interaction. Let us come back to formula (\ref{exp}).
Again we put in it
\[ \cosh(\phi^\prime/2)\cosh(\phi/2)
-(\n^\prime \n)\sinh(\phi^\prime/2)\sinh(\phi/2)\,=\,1. \]
In the other terms, linear in $\si$, we keep only the first power of
$\q\rightarrow - i\hbar \nabla$, in the hope that in the final
result (\ref{prod}) $\hbar$ will enter in the
combination $\hbar s\rightarrow$ const. Nevertheless, these
terms by themselves are small as compared to unity, so that in
the classical limit expression (\ref{exp}) can be rewritten as
\[ \exp\left\{
\si\left[\n^\prime\sinh(\phi^\prime/2)\cosh(\phi/2)
-\n\cosh(\phi^\prime/2)\sinh(\phi/2)\right]\right. \]
\[ \left.-\,i\,(\si[\n^\prime \times \n])\sinh^2(\phi/2)\right\}. \]
Clearly, in the product (\ref{prod}) the
operators $\si$ attached to
\[ \n^\prime\sinh(\phi^\prime/2)\cosh(\phi/2)
-\n\cosh(\phi^\prime/2)\sinh(\phi/2)\,,\]
combine in the resulting exponent into the operator $\Si$
which vanishes in the classical limit. In this limit only
those operators $\si$ survive that are attached to
$\;[\n^\prime \times \n]\sinh^2(\phi/2);\;$
they combine into $2\s$. Thus, in the classical limit the
product (\ref{prod}) reduces, with the account for the second
identity (\ref{ide2}), to
\beq\label{Prod}
\exp\left\{ {1 \over m}{\gamma \over \gamma+1}\,\left(\s\,[\bv\times
\nabla]\right)\right\}.
\eeq

Let us note that the action of the operator (\ref{Prod}) on
any function of coordinates, whether it is a vector potential or
field strength, amounts to the shift of its argument:
\[ \r \rightarrow \r + {1 \over m}\,{\gamma \over \gamma+1}\,
\s \times \bv\,. \]
Curiously, just this substitution was pointed out in~\cite{hei}
for the transition from covariant equations linear in spin
to noncovariant equations (see (\ref{defh})).

Now, taking into account the second term in the expansion of the
exponential function (\ref{Prod}) and using again the identity
(\ref{ide}), we obtain the following expression for the
quadratic in spin interaction arising from the convection
current:
\[-\,{e \over 2m^2}\,{\gamma \over \gamma+1}\,\left(\s\,[\bv\times
\nabla]\right)\left[\left(1-\,{1 \over \gamma}\right)(\s\B)\right.\]
\beq\label{s2c}
\left. -\,{\gamma \over \gamma+1}\,(\s\bv)(\bv\B)\,-\,
{\gamma \over \gamma+1}\left(\s\,[\bv\times \E]\right)\right].
\eeq

Let us discuss now the contribution to the discussed
effect due to the magnetic moment. It is convenient to write the 
term in the scattering amplitude we are interested in (it is 
proportional to $g$-factor) as
\[\frac{eg}{4\ep}\;\xi_0^{\prime \dagger}
\left\{[\exp\{\Si\fib^\prime/2\}\exp\{-\Si\fib/2\}
\exp\{\Si\fib/2\}(\s\B)\exp\{-\Si\fib/2\}\right. \]
\[ +\exp\{-\Si\fib^\prime/2\}\exp\{\Si\fib/2\}
\exp\{-\Si\fib/2\}(\s\B)\exp\{\Si\fib/2\}] \]
\beq\label{lseg2}
+\,{i \over 2}\,[\exp\{\Si\fib^\prime/2\}\exp\{-\Si\fib/2\}
\exp\{\Si\fib/2\}(\Si\E)\exp\{-\Si\fib/2\}
\eeq
\[ \left. -\exp\{-\Si\fib^\prime/2\}\exp\{\Si\fib/2\}
\exp\{-\Si\fib/2\}(\Si\E)\exp\{\Si\fib/2\}]\right\}\xi_0. \]
Using in this case the first term in the expansion of the
exponential function (\ref{Prod}), we arrive at the following
expression for the contribution proportional to the magnetic
moment:
\beq\label{s2g}
\,{eg\over 2m^2}\,{\gamma \over \gamma+1}\,\left(\s\,[\bv\times
\nabla]\right)\left[(\s\B)\,-\,
{\gamma \over \gamma+1}\,(\s\bv)(\bv\B)\,
-\,\left(\s\,[\bv\times \E]\right)\right].
\eeq
The total result for the induced interaction, quadratic in the spin,
is
\[ L_{2s}^i=\,{e \over 2m^2}\,{\gamma \over \gamma+1}
\,\left(\s\,[\bv\times \nabla]\right)
\left[\left(g-1+\,{1 \over \gamma}\right)(\s\B)\,-\,
(g-1)\,{\gamma \over \gamma+1}\,(\s\bv)(\bv\B)\,\right. \]
\beq\label{qu2}
\left. -\,\left(g-{\gamma \over \gamma+1}\right)
\left(\s\,[\bv\times \E]\right)\right].
\eeq

Let us note that in the nonrelativistic limit the induced
interaction with magnetic field tends to zero as $v/c$, and
that with electric field as $(v/c)^2$. Moreover, the part of
interaction (\ref{qu2}) that is not related to $g$-factor,
is reducible in spin; in other words,  
$\,s_i s_j\,$ in it cannot be rewritten as an irreducible tensor
$\,s_i s_j - (1/3)\delta_{ij}\s^{\,2}\,$. Therefore, the interaction 
(\ref{qu2}) is not in fact a quadrupole one. However, its 
asymptotic behaviour for $\,\gamma\rightarrow \infty\,$ is of 
interest. In this limit
\beq\label{quasi}
L_{2s}^i=\,{e \over 2m^2}\,(g-1)\,\left(\s\,[\bv\times \nabla]\right)
\,[(\s\B)\,-\,(\s\bv)(\bv\B)\,-\,
\left(\s\,[\bv\times \E]\right)].
\eeq
Surprisingly, the asymptotical formulae (\ref{quas}) and
(\ref{quasi}) coincide to within a factor and a total time 
derivative. To prove this, it is convenient to introduce three
orthogonal unit vectors
\[ \bv;\;\;\vrho\,=\,{[\bv\times \s] \over |[\bv\times \s]|};\;\;
   \vzeta\,=\,[\bv\times \vrho]. \]
Using the completeness of this basis and the equation $\;\dot{\E}=
[\nabla\times \B]\,,$ and discarding a total derivative with respect
to $\,t\,$, one can check that
\[ [(\s\nabla)\,-\,(\bv\s)(\bv \nabla)]\,
[(\s\E)\,-\,(\s\bv)(\bv\E)\,+\,(\s[\bv\times\B])] \]
\[ =\,[\bv\times \s]^2\,(\vzeta\nabla)[(\vzeta\E)\,+\,(\vrho\B)]\;, \]
coincides indeed with
\[ \left(\s\,[\bv\times \nabla]\right)
\,[(\s\B)\,-\,(\s\bv)(\bv\B)\,
-\,\left(\s\,[\bv\times \E]\right)] \]
\[ =\,-\,[\bv\times \s]^2\,(\vrho\nabla)
\left[\left(\vrho[\bv\times \B]\right)\,+\,(\vrho\E)\right]. \]
Thus, there is a special value of the bare quadrupole moment
\beq\label{quab}
Q\,=\,-\,2(g-1)\,{es^2 \over m^2}\,,\;\;\;\mbox{or}\;\;
f_{e,2}\,=\,(g-1)\,{1 \over 2m^2}
\eeq
(let us recall that we consider now a classical situation, when
$\,s\gg 1\,$), at which the total interaction quadratic in
spin, $\,L_{2s}+L_{2s}^i\,$, asymptotically decreases with
energy.

The situation resembles that which takes place for the
interaction linear in spin. It is well-known (see, for 
instance,~\cite{kh,wei,fpt}) that there is a special value of 
$g$-factor,
$\,g=2\,$, at which the interaction linear in spin decreases
as $\;\gamma\rightarrow \infty\,$. This follows immediately from
formula (\ref{lse}) for the first-order Lagrangian. Thus,
putting additionally $\,g=2\,$, we obtain
\beq\label{quab1}
Q\,=\,-\,2\,{es^2 \over m^2}\,,\;\;\;\mbox{or}\;\;
f_{e,2}\,=\,{1 \over 2m^2}\,.
\eeq

Let us note that the choice $\,g=2\,$ for the bare magnetic
moment is a necessary (but insufficient) condition of
renormalizability in quantum electrodynamics \cite{kh,wei,fpt}.
It is satisfied not only for the electron, but also for the charged
vector boson in the renormalizable electroweak theory.

In some respect, however, the situation with the special
values (\ref{quab}), (\ref{quab1}) of the quadrupole moment
differs from the situation with $g$-factor. The conditions
(\ref{quab}), (\ref{quab1}), as distinct from the condition
$\,g=2\,$, are not universal, since they are valid only for large
spins, $\;s\gg 1\,$; in other words, they refer only to
classical objects with internal angular momentum. In
particular, for the charged vector boson of the renormalizable
electroweak theory the bare quadrupole interaction is absent
at all, $\;f_{e,2}=0\,$. The quadrupole moment of this particle
is (in our language) of the induced nature, it is given by
formula (\ref{qua1}) at $\,s=1\,$ ¨ $\,g=2\,$.

\section{A Simple-Minded Aside on the Spin\\ Precession in a 
Gravitational Field}
In this section we present a simple and general derivation of
the equations of spin precession in a gravitational field.
This approach not only allows us to easily reproduce and
generalize known results for spin effects. Pointed out here
remarkable analogy between gravitational and electromagnetic
fields allows also to easily transform the results of the
previous section to the case of an external gravitational
field.

It follows from the angular momentum conservation in flat
space-time taken together with the equivalence principle that
the 4-vector of spin $S^\mu$ is parallel transported along the
particle world-line. The parallel transport of a vector
along a geodesic $x^\mu(\tau)$ means that its covariant
derivative vanishes:
\beq\label{par}
\frac{DS^\mu}{D\tau}=\,0\,.
\eeq
(In this section we restrict our discussion to the effects linear 
in the spin.)  We will use the tetrad formalism natural for 
the description of spin. In view of relation (\ref{par}), the
equation for the tetrad components of spin
$S^a=\,S^\mu e^a_{\mu}$ is
\begin{equation}\label{pars}
\frac{DS^a}{D\tau}=\,\frac{dS^a}{d\tau}=\,S^\mu e^a_{\mu;\nu}u^\nu=
\,\eta^{ab}\gamma_{bcd}u^d S^c\,.
\end{equation}
Here
\beq\label{rota}
\gamma_{abc}=\,e_{a\mu;\nu}e^\mu_{b}e^\nu_{c}=\,-\gamma_{bac}
\eeq
are the Ricci rotation coefficients~\cite{ll}. Certainly, the
equation for the tetrad 4-velocity components is exactly the
same:
\begin{equation}\label{paru}
\frac{du^a}{d\tau}=\,\eta^{ab}\gamma_{bcd}u^d u^c\,.
\end{equation}
The meaning of Eqs.~(\ref{pars}), (\ref{paru}) is clear: the
tetrad components of both vectors vary in the same way, due to
the rotation of the local Lorentz vierbein only.

In exactly the same way, the 4-dimensional spin and velocity of
a charged
particle with the gyromagnetic ratio $g=2$ precess with the same
angular velocity in an external electromagnetic field, by virtue
of equation (\ref{bmt0}) at $g=2$ and the Lorentz equation:
\[ \frac{dS_a}{d\tau}=\,\frac{e}{m}F_{ab}S^b;\;\;\;\;\;
\frac{du_a}{d\tau}=\,\frac{e}{m}F_{ab}u^{b}.\]

Thus, the correspondence:
\beq\label{corcov}
\frac{e}{m}F_{ab} \longleftrightarrow \gamma_{abc}u^c.
\eeq
gets obvious. This correspondence allows one to obtain the 
precession frequency $\vom$ of the
3-dimensional vector of spin $\s$ in external gravitational field 
from expression (\ref{Om}) via the simple substitution
\beq\label{cornon}
\frac{e}{m}B_i \longrightarrow
-\,\frac{1}{2}\epsilon_{ikl}\gamma_{klc}u^c;
\;\;\; \frac{e}{m}E_i \longrightarrow \gamma_{0ic}u^c.
\eeq
This frequency is
\begin{equation}\label{og1}
\omega_i=\,-\epsilon_{ikl}\left(\frac{1}{2}\gamma_{klc}+
\,\frac{u^k}{u^0+1}\gamma_{0lc}\right)\,\frac{u^c}{u^0_w}\,.
\end{equation}
The factor $1/u^0_w$ in this expression is related to the
transition in the left-hand side of Eq.~(\ref{pars}) to the 
differentiation with respect to the world time $t$:
$$\frac{d}{d\tau}=\,\frac{dt}{d\tau}\,\frac{d}{dt}
=\,u^0_w\,\frac{d}{dt}.$$
The quantity $u^0_w$ is supplied with the subscript $w$ to
emphasize that this is a world component of 4-velocity, but not a
tetrad component. All other indices in expression (\ref{og1}) are 
tetrad ones, $c=0,1,2,3;\;\;i,k,l=1,2,3$.
The corresponding spin-dependent correction to the Lagrangian is
\begin{equation}\label{sg1}
L_{1sg}=\,\s \vom\,.
\end{equation}

As an illustration of formulae (\ref{og1}), (\ref{sg1}), let us
apply them to the cases of spin-orbit and spin-spin interactions.
We restrict, as is common in the problems discussed, to the
linear approximation in the gravitational field. However, in our
approach, as distinct from the standard ones, both problems can be
easily solved for arbitrary particle velocities.

The tetrads $e_{a\mu}$ are related to the metric as follows:
$$e_{a\mu}e_{b\nu}\eta^{ab}=\,g_{\mu\nu}.$$
To linear approximation we can put
$g_{\mu\nu}=\,\eta_{\mu\nu}+\,h_{\mu\nu}$ and do not distinguish
between the tetrad and world indices in $e_{a\mu}$.
The ambiguity in the choice of tetrads will be fixed by choosing
the symmetric gauge $e_{\mu\nu}=\,e_{\nu\mu}\,.$ Then
$$e_{\mu\nu}=\,\eta_{\mu\nu}+\,\frac{1}{2}h_{\mu\nu}\,.$$
Using expression (\ref{rota}) for the Ricci coefficients, we
find to linear approximation
\begin{equation}\label{gam}
\gamma_{abc}=\,\frac{1}{2}(h_{bc,a}-\,h_{ac,b})\,.
\end{equation}

Let us start with the spin-orbit interaction. In the centrally
symmetric field created by a mass $M$, the metric is
\begin{equation}\label{lin}
h_{00}=\,-\frac{2kM}{r};\;\;\; h_{mn}=\,-\frac{2kM}{r}\delta_{mn}.
\end{equation}
Here the nonvanishing Ricci coefficients are
\begin{equation}\label{ric}
\gamma_{ijk}=\,\frac{kM}{r^3}(\delta_{jk}r_i-\,\delta_{ik}r_j)\,,
\;\;\;\;\;\gamma_{0i0}=\,-\frac{kM}{r^3}r_i\,.
\end{equation}
Their substitution into formula (\ref{og1}) yields the following
expression for the precession frequency:
\begin{equation}\label{so}
{\vom_{ls}}=\,\frac{2\gamma+\,1}{\gamma+\,1}\,\frac{kM}{r^3}\,
\bv \times \r \,.
\end{equation}
In the limit of low velocities, $\gamma \rightarrow 1$, the
answer goes over into the classical result~\cite{des}.

Now we consider the spin-spin interaction. Let the spin of
the central body be $\s _0$. Linear in $\s _0$ components of
metric, which are responsible for the spin-spin interaction,
are:
$$h_{0i}=\,2k\,\frac{[\s _0 \times \r]_i}{r^3}\,.$$
Here the nonvanishing Ricci coefficients are
\begin{equation}\label{ri1}
\gamma_{ij0}=\,k\left(\nabla_i\frac{[\s _0\times\r]_j}{r^3}-\,
\nabla_j\frac{[\s _0\times \r]_i}{r^3}\right)\,,\;\;\;\;\;
\gamma_{0ij}=\,-k\nabla_i\frac{[\s _0 \times \r]_j}{r^3}\,.
\end{equation}
The frequency of the spin-spin precession is  
$$ \vom_{ss}=\,-k\,\left(2-\frac{1}{\gamma}\right)
(\s _0\nabla)\nabla\frac{1}{r}$$
\begin{equation}\label{ss}
+\,k\,\frac{\gamma}{\gamma+\,1}
\left[\bv (\s _0\nabla) -\,\s _0(\bv \nabla)
+\,(\bv \s _0)\nabla\right]\,(\bv \nabla)\,\frac{1}{r}\,.
\end{equation}
In the limit of low velocities this formula also goes over into
the corresponding classical result~\cite{sch}.

In the conclusion of this section we note that in the case of
an external gravitational field there is no covariant
expression for the force linear in the particle spin. In other
words, the deviation from geodesics of the trajectory of a
spinning particle is not described by the Riemann tensor. If it 
were the case, there would be a unique possible covariant 
structure, to within a factor (in~\cite{pa} it equals $-\,1/2m$):
$R_{\mu\nu ab}u^{\nu}S^{ab}$. As mentioned already in Section 1, 
the covariant description (as distinct from our
our formulae (\ref{og1}), (\ref{sg1})) contradicts the classical
results in the limit of low velocities.

\section{Equations of Motion of a Spinning Particle\\
         in a Gravitational Field}
\subsection{The General Approach}
The equations of motion in an external gravitational field to any
order in spin are constructed similarly to the equations of
motion in the case of an electromagnetic field.

We start with the elastic scattering amplitude in a weak 
external gravitational field $h_{\mu\nu}$. We use it as an euristic
argument only, and afterwards will go beyond the linear 
approximation. This amplitude is
\beq\label{gampl}
-\,{1 \over 2}T_{\mu\nu}h^{\mu\nu}.
\eeq
The matrix element $T_{\mu\nu}$ of the energy-momentum tensor
between states of momenta $k$ and $k^\prime$ can be written
as
\[ T_{\mu\nu}=\frac{1}{4\ep}\,\bar{\psi}(k^\prime)
\left\{p_{\mu}p_{\nu}\,F_1
 +\,{1 \over 2}\,(p_{\mu}\Sigma_{\nu\lambda}
+\,p_{\nu}\Sigma_{\mu\lambda})\,q^{\lambda}\,F_2 \right. \]
\beq\label{tens}
 +\,(\eta_{\mu\nu}q^2\,-\,q_\mu q_\nu)\,F_3
\eeq
\[ \left. +\,[S_\mu S_\nu q^2 -\,(S_\mu q_\nu+\,S_\nu q_\mu)(Sq)\,+\,
\eta_{\mu\nu}(Sq)^2]\,F_4 \right\}\,\psi(k). \]

The scalar operators $F_i$ in this expression are also expanded
in powers of $\tau=(Sq)^2$:
\[ F_i(t,\tau)=\sum_{n=0}^{N_i}f_{i,2n}(t)\tau^n. \]
It can be easily proven that the total number of invariant
form factors $f_{i,2n}$ is $4s+2$ and $4s+1$ for integer and
half-integer spin, respectively. The independence of the four
tensor structures in (\ref{tens}) is obvious. As
for the completeness of the expansion, it can be proven, for
instance, by demonstrating that the total number of invariant form 
factors, as calculated in the annihilation channel, coincides with 
the above result.          

In the generally covariant form, the structure
$(\eta_{\mu\nu}q^2\,-\,q_\mu q_\nu)\,h^{\mu\nu}$ corresponds to
the scalar curvature $R$, and
$[S_\mu S_\nu q^2 -\,(S_\mu q_\nu+\,S_\nu q_\mu)(Sq)\,+\,
\eta_{\mu\nu}(Sq)^2]\,h^{\mu\nu}$ corresponds to the product
$R_{\mu\nu} S^\mu S^\nu$, where $R_{\mu\nu}$ is the Ricci tensor.
Since we are interested in the equations of motion in a
sourceless field, the corresponding terms in the expansion
(\ref{tens}) will be omitted.

Just as in electrodynamics charge conservation $\;$ dictates $\;$
the $\;$ condition $\,f_{e,0}(0)=1\,$, here energy
conservation leads to $\,f_{1,0}(0)=1\,$. As to the term in the
amplitude (\ref{gampl}) which contains $\,f_{2,0}\,$, it is
convenient to write it in a different form, using the analogy
(\ref{corcov}) with electromagnetic field. Putting $g=2$ and
$(e/m)F_{ab}\rightarrow f_{ab}=\gamma_{abc}u^c\,$ in the
corresponding electromagnetic term
\[ i\,\frac{eg}{8\ep}\,\bar{\psi}(k^\prime)\,\Sigma^{ab}
F_{ab}\,\psi(k), \]
we arrive at the following contribution to the Lagrangian of 
the gravitational interaction:
\beq\label{gamp2}
i\,\frac{1}{4u^0_w}\,\bar{\psi}(k^\prime)\,\Sigma^{ab}f_{ab}\,
\psi(k);
\eeq
here, as usual, $\;u_w^0=\,\ep/m$. Using for
$\,\gamma_{abc}\;$ the linear approximation (\ref{gam}), one can
easily see that the expression (\ref{gamp2}) corresponds indeed
to the discussed contribution to the amplitude provided that 
$\;f_{2,0}=1$. Thus, in gravity the value of one more form
factor at zero momentum transfer $t$ is fixed. It corresponds to
conservation of angular momentum. This was pointed out earlier
in~\cite{kook,heh}.

Let us now come back to the convection term in formula 
(\ref{gampl}).
As in electrodynamics, when we go over here to
spinors in the rest frame, the term of first order in spin is
written as
\beq\label{gcon}
-\,{p^{\mu}p^{\nu} \over 8 \ep}\,{1 \over m}\,{u^0 \over u^0+1}
\left(\s\,[\bv\times \nabla]\right)\,h_{\mu\nu}.
\eeq
Using (\ref{ide}), (\ref{gam}), we obtain
\[ p^\mu \nabla_k h_{\mu\nu}\rightarrow
-\,p^\mu (-\partial_k h_{\mu\nu}+\,\partial_\mu h_{k\nu})
\rightarrow -\,2 p^a \gamma_{ak\nu}.\]
Thus, the expression (\ref{gcon}) can be written in terms of the 
Ricci coefficients:
\beq\label{gcon1}
{1 \over u^0_w}\,{u^0 \over u^0+1}\,\ep^{mnk} s^m v^n
u^a u^c \gamma_{akc}.
\eeq
As can be easily seen, the sum of (\ref{gamp2}) and
(\ref{gcon1}) reproduces the Lagrangian (\ref{sg1}).

\subsection{Effects Quadratic in the Spin}

Let us investigate now the effects of second order in the spin in
the equations of motion in a gravitational field. In the case  
of a binary star these effects are of the same order of magnitude
as the spin-spin interaction when the spins of the components of
the system are comparable~\cite{khp}. The influence of the latter 
on the characteristics of gravitational radiation becomes noticeable
for a system of two extreme black holes~\cite{kww}. Correspondingly,  
second-order spin effects in the equations of motion become
substantial if at least one component of a binary is close to an 
extreme
black hole~\cite{khp}. Therefore, the investigation of these
effects is not of a purely theoretical interest only. In
principle they can be observed with the gravitational wave
detectors under construction.

An obvious source of second-order spin effects is the term
\beq\label{agm}
L_{2sg}=\,-\,f_{1,2}\,{1 \over 8\ep}\,p^\mu p^\nu (Sq)^2 h_{\mu\nu}
\eeq
in the amplitude (\ref{gampl}). Due to the relation
\[ p^\mu p^\nu q_\alpha q_\beta h_{\mu\nu}=\,
 p^\mu p^\nu ( q_\alpha q_\beta h_{\mu\nu}+
 \,q_\mu q_\nu h_{\alpha\beta}-\,
 q_\alpha q_\nu h_{\mu\beta}-\,q_\beta q_\mu h_{\nu\alpha}) \]
\[ \rightarrow 2 p^\mu p^\nu R_{\mu\alpha\nu\beta}, \]
the Lagrangian (\ref{agm}) can be written in terms of the Riemann 
tensor:
\beq\label{agmr}
L_{2sg}=\,-\,{\kappa  \over 2\ep }\,u^a S^b u^c S^d R_{abcd}.
\eeq
Instead of $\,f_{1,2}\,$, we have introduced a dimensionless
parameter $\kappa$:
\[ f_{1,2}=\,{\kappa \over 2m^2}. \]

Now, it is convenient to use the Petrov representation for the
components of the Riemann tesor (see~\cite{ll}):
\[ E_{kl}=\,R_{0k0l},\;\;E_{kl}=\,E_{lk};\;\;\;
   C_{kl}=\,\frac{1}{4}\epsilon_{kmn}\epsilon_{lrs}R_{mnrs},\;\;
   C_{kl}=C_{lk}; \]
\beq\label{pe1}
B_{kl}=\,\frac{1}{2}\epsilon_{lrs}R_{0krs},\;\;B_{kk}=\,0.
\eeq
We limit our discussion to the case of a sourceless gravitational 
field.
Then, at $R_{ab}=0$, further simplifications hold:
\beq\label{pe2}
C_{kl}=\,-\,E_{kl},\;\;\;B_{kl}=\,B_{lk},
\;\;\;E_{kk}=\,C_{kk}=\,0.
\eeq
Finally, we arrive at the following interaction Lagrangian
quadratic in the spin:
\[ L_{2sg}=\,-\,{\kappa \over 2\ep}\,
\left[(2\bu^2+1)\,E_{kl}-\,2\,\left(2
-\,{1 \over u_0+\,1}\right) u_k u_m E_{lm}
+\,\delta_{kl}u_m u_n E_{mn}\right. \]
\beq\label{sg2}
+\,{1 \over (u_0+\,1)^2}\,u_k u_l u_m u_n E_{mn}
\eeq
\[ \left. -\,2\,u_0\,\epsilon_{kmn}u_m B_{nl}
+\,{2 \over u_0+\,1}
\,u_k u_m \epsilon_{lrn} u_r B_{mn}\right]\,
(s_k s_l - {1 \over 3}\,\delta_{kl} \s ^2). \]
To avoid misunderstandings, we note that all three-dimensional
indices in this equation (and in (\ref{sg2i}), (\ref{g2tot})) are 
in fact contravariant.

As in electrodynamics, along with the ``bare'' interaction 
(\ref{sg2}), there is here an induced interaction quadratic in
spin. Its explicit form can be obtained most easily by setting
$g=2$ in the electromagnetic formula (\ref{qu2}) and by making the
substitution (\ref{cornon}). We also take into account the
correspondence
\[ q_i \gamma_{abc}u^c\,=\,(q_i \gamma_{abc}
-\,q_c \gamma_{abi})\,u^c\,\rightarrow
\,i\,(\partial_i \gamma_{abc}
-\,\partial_c \gamma_{abi})\,u^c\,\rightarrow
\,i\,R_{abci}u^c. \]
Finally, using (\ref{pe1}) and (\ref{pe2}), we obtain the
following result for the induced interaction:
\[ L_{2sg}^i\,=\,{1 \over 2\ep}\,
\left\{\left(2\bu^2-\,{u^0-1 \over u^0+\,1}\right)\,E_{kl}
-\,2\,\left[2
-\,{1 \over u_0+\,1}-\,{1 \over (u_0+\,1)^2}\right]\,
 u_k u_m E_{lm}\right. \]
\beq\label{sg2i}
+\,\left[1-\,{1 \over (u_0+\,1)^2}\right]\,
\delta_{kl}u_m u_n E_{mn}
+\,{1 \over (u_0+\,1)^2}\,u_k u_l u_m u_n E_{mn}
\eeq
\[ \left. -\,2\,\left(u_0-\,{1 \over u_0+\,1}\right)
\epsilon_{kmn}u_m B_{nl}
+\,{2 \over u_0+\,1}
\,u_k u_m \epsilon_{lrn} u_r B_{mn}\right\}\,s_k s_l. \]
As in the electromagnetic case, the induced interaction
tends to zero in the nonrelativistic limit  $\,\sim v/c\,$,
and the spin factor in it, $\,s_k s_l\,$, is not an irreducible
tensor.

The asymptotic behaviour of $L_{2sg}$ and $L_{2sg}^i$ is the
same: both Lagrangians increase linearly with energy.
However, in this case too the coefficient in the
``bare'' interaction can be chosen in such a way, $\,\kappa=1\,$,
that the total Lagrangian of second order in the spin decreases
(as well as the analogous interaction in electrodynamics)
when the energy tends to infinity. At $\,\kappa=1\,$
\[ L_{2sg}+\,L_{2sg}^i\,=\,-\,{1 \over \ep (u^0+\,1)}\,
\left(u^0\,E_{kl}-\,{1 \over u_0+\,1}\,u_k u_m E_{lm}\right. \]
\beq\label{g2tot}
\left. +\,{1 \over 2(u_0+\,1)}\,\delta_{kl}u_m u_n E_{mn}
+\,\epsilon_{kmn}u_m B_{nl} \right)\,s_k s_l.
\eeq

\section{The Gravimagnetic Moment}

There is a profound analogy between the linear in the spin 
interaction of the magnetic moment with the electromagnetic field
\beq\label{Le}
{\cal L} _{em}=-\frac{eg}{4m}F_{ab}S^{ab}\,
\eeq
and the ``bare'' gravitational Lagrangian (\ref{agmr}) which is 
quadratic in the spin~\cite{kh}. (Here it is more convenient to 
write the
gravitational Lagrangian, like $\,{\cal L} _{em}\,$, for the
proper time $\,\tau\,$, rather than the world time $\,t\,$, i.e.,
to multiply expression (\ref{agmr}) by $\,\ep/m\,$.) This
analogy is based on the following observation~\cite{kh}. It is 
well-known
that the canonical momentum $\,i\partial_\mu\,$ enters
relativistic wave equations for a particle in electromagnetic and
gravitational external fields via the combination
\[ \Pi_\mu = \,i\partial_\mu - eA_\mu -
{1 \over 2}\Sigma^{ab}\gamma_{ab\mu}. \]
It follows from the structure of the commutator (or Poisson
brackets in the classical limit)
\[ [\Pi_\mu,\Pi_\nu]=\,-\,i\,(eF_{\mu\nu}
-\,{1 \over 2}\Sigma^{ab}R_{ab\mu\nu}) \]
that in a sense $\,-\,{1 \over 2}
\Sigma^{ab}R_{ab\mu\nu}\,$ plays the same role in gravity as
$\,eF_{\mu\nu}\,$ in electromagnetism. It is quite natural then
that the gravitational analogue of the electromagnetic spin
interaction (\ref{Le}) is
\begin{equation}\label{gm}
{\cal L} _{gm}=\,\frac{\kappa}{8m}R_{abcd}S^{ab}S^{cd}\,.
\end{equation}
One can easily show that expressions (\ref{gm}) and
(\ref{agmr}) coincide (to within a factor $\,\ep/m\,$). It is
sufficient to this end to take into account the relation
$\,S^{ab}=\,\ep^{abcd}S_c u_d\,$, as well as the identity
\[ \tilde{R}_{abcd}=\,{1 \over 4}
\,\ep_{ab}^{\;\;\;ef}\ep_{cd}^{\;\;\;gh}
R_{efgh}=\,-\,R_{abcd}\, , \]
which is valid for a sourceless gravitational field.

By analogy with the magnetic moment
$$\frac{eg}{2m}S^{\mu\nu}\,,$$
it is natural to define the gravimagnetic moment
$$-\,\frac{\kappa}{2m}S^{ab}S^{cd}\,.$$
The gravimagnetic ratio $\kappa$, like the gyromagnetic ratio
$g$ in electrodynamics, may have in general any value. However,
according to our semiclassical arguments, the value $\kappa=1$ 
in gravity is as preferable as $g=2$ in electrodynamics. In any 
case, at $g=2$ and $\kappa=1$ the spin equations of motion have 
the simplest form.

On the other hand, it has been shown in~\cite{kh} that just this 
value of the gravimagnetic ratio, $\kappa=1$, follows from the 
wave equations in the Feynman gauge for the photon and graviton
in an external gravitational field. The same value, $\kappa=1$,
follows from the squared Rarita-Schwinger equation for $s=3/2$ in a 
gravitational field~\cite{kh}.

A second-order wave equation for an arbitrary spin in a 
gravitational background had been proposed long ago in~\cite{chd}.
For integer spins, its form corresponds also to $\kappa=1$. However,
the value of $\kappa$ proposed in~\cite{chd} for half-integer
spins is quite different, it does not tend to unity even in the 
semiclassical limit $s\rightarrow\infty$. Clearly, such a 
prescription does not look reasonable: in the semiclassical limit 
$s\rightarrow\infty$ the parameter $\kappa$ should not change when 
the spin changes by 1/2.

The situation with spin 1/2 is rather special~\cite{kh}. 
The properties of the spin matrices for $s=1/2$,
\[ S^{ab}={i \over 4}(\gamma^a \gamma^b - \gamma^b \gamma^a), \]
are such that the gravimagnetic interaction (\ref{gm})
degenerates here into the mere scalar curvature
\beq\label{1/8}
{\kappa \over 16 m}R,
\eeq 
without any consequences whatsoever for the motion of the spin.
Obviously, the coefficient at this spin-independent stucture 
cannot be fixed by the above arguments. Formula (\ref{1/8}) 
would predict for $\kappa=1$ the following wave equation for 
$s=1/2$
\beq\label{deq0}
\left(-g^{\mu\nu}D_{\mu}D_{\nu}-m^2+{1 \over 8}R\right)\psi=0
\eeq
(the factor $1/2m$ should be deleted from the expressions 
(\ref{gm}) and (\ref{1/8}) when going over from a Lagrangian in our 
normalization to a wave equation). Meanwhile, the squared Dirac 
equation in a gravitational field is
\beq\label{deq}
\left(-g^{\mu\nu}D_{\mu}D_{\nu}-m^2+{1 \over 4}R\right)\psi=0.
\eeq
The discrepancy between (\ref{deq0}) and (\ref{deq}) can be removed,
for instance, by adding 
\[ {1 \over 16m}R \]
to the Lagrangian (\ref{gm}) for half-integer spins. This term in 
no way influences our semiclassical
arguments, vanishing in the limit $\hbar\rightarrow 0$.
As to the recent proposal~\cite{aap} to cure the discussed 
discrepancy by ascribing to the electron (which has no 
gravimagnetic interaction at all) the gravimagnetic ratio 
$\kappa=2$, we cannot see in it any real physical meaning.

\section{Multipoles of Black Holes}
But let us come back from elementary particles to macroscopic 
bodies. For a classical object the values of both parameters $g$ and
$\kappa$ depend in general on the various properties of the body. 
However,
for black holes the situation is different. It has been shown 
in~\cite{cart} from an
analysis of the Kerr-Newman solution that the gyromagnetic ratio
of a charged rotating black hole is universal (and equal to 
that of the electron!): $g=2$.

We will show that for the Kerr black hole the gravimagnetic
ratio is $\kappa=\,1$. This value follows in fact from the
analysis of the motion of spin of a black hole in an external
field done in~\cite{hath} (though this statement was not 
explicitly formulated there). We will present here
an independent and, in our opinion, simpler derivation of this
important result.

At great distance from a Kerr hole, the hole can be considered as a
point source of a weak gravitational field. To linear 
approximation in the field of a hole at rest, the Lagrangian 
density corresponding to the interaction (\ref{agmr}) can be
written as
\begin{equation}\label{den}
\L=\,\frac{\kappa}{4m}\,(\s \nabla)^2\, h_{00} \delta(\r)\,.
\end{equation}
The thus induced correction to the energy-momentum tensor has
a single component:
\begin{equation}\label{ten}
\delta T_{00}=\,-\,\frac{\kappa}{2m}\,(\s \nabla)^2\, \delta(\r)\,.
\end{equation}
In the gauge
\begin{equation}\label{cal}
\bar{h}^{\mu\nu},_{\nu}=\,0,\;\;
\bar{h}_{\mu\nu}=\,h_{\mu\nu}
-\,\frac{1}{2}\eta_{\mu\nu}h^\alpha_\alpha
\end{equation}
the static Einstein equation for the corresponding correction 
$h_{00}$ to the 00-component of the metric is 
$$\Delta h_{00}=\,8 \pi k T_{00}\,.$$
The correction itself is
\beq\label{h00}
h_{00}=\,\kappa \frac{k}{m}\,(\s \nabla)^2\,\frac{1}{r}\,.
\eeq
 
Let us compare $h_{00}$ with the corresponding contribution to the 
Kerr metric. In the Boyer-Lindquist coordinates this metric is
\[ 
ds^2=\,(1-\,\frac{r_g r}{\Sigma})dt^2-\,\frac{\Sigma}{\Delta}dr^2-\,
\Sigma d\theta^2-\,(r^2+\,a^2+\,\frac{r_g ra^2}{\Sigma}\sin^2 \theta)
r^2 \sin^2 \theta \]
\begin{equation}\label{ker}
+\,\frac{2r_g ra}{\Sigma}\sin^2 \theta d\phi dt\,,
\end{equation}
where $\Delta=\,r^2-\,r_g r+\,a^2\,,\;\;
\Sigma=\,r^2+\,a^2 \cos^2 \theta\,,\;\;\va=\,\s/m.$
At $r_g=\,0$ the metric (\ref{ker}) describes a flat space in
spheroidal coordinates \cite{ll}. Meanwhile, it is Cartesian
coordinates which correspond in the flat space to the gauge
(\ref{cal}). The transition from the spheroidal coordinates to
Cartesian ones is carried out with the required accuracy by the
substitution
$$\r\rightarrow\,\r+\,\frac{\va(\va \r)-\,\r a^2}{2r^2}\,.$$
In the Cartesian coordinates the spin-dependent part of the
00-component of the metric
$$ g_{00}=\,1-\,\frac{r_g}{r}+\frac{r_g a^2}{2r^3}\,
(3\cos^2 \theta-\,1)\,$$
obviously coincides with $h_{00}$ from formula (\ref{h00}) at
$\kappa =\,1$. Somewhat more tedious consideration of the space 
components of the Kerr metric leads to the same result, 
$\kappa =\,1$.

Let us note that the motion of the Kerr black hole in an external
gravitational field is not described by the Papapetrou equation
even if one leaves aside the problem of spin-orbit interaction
linear in spin. The point is that this equation refers to the case
$\kappa=\,0\;$ \cite{yb}.

It is proven in the same way that for a charged Kerr hole as well
the gravimagnetic ratio $\kappa=\,1$. Moreover, it can be proven
that the electric quadrupole moment of a charged Kerr hole also
equals
\[ \,Q\,=\,-\,2\,{es^2 \over m^2}\,, \]
the value, at which the interaction quadratic in spin decreases 
with energy. It has been shown~\cite{pom} that other, higher 
multipoles of a charged Kerr hole as well possess just those values 
which guarantee that the interaction of any order in spin (but of 
course, linear in an external field) asymptotically decreases with 
increasing energy.

\section*{Acknowledgements}
We are grateful to I.A. Koop, R.A. Sen'kov, A.N. Skrinsky, and
Yu.M. Shatu- nov for useful discussions. We are also grateful to 
D. Bini, G. Gemelli, and R. Ruffini for the advices due to which, 
hopefully, the present text has become more comprehensible.
We greatly appreciate the warm hospitality extended to us by ICRA. 
The work was supported by the Russian Foundation for Basic Research 
through Grant No. 98-02-17797 and by the Federal Program 
Integration-1998 through Project No. 274.

\end{document}